\documentclass[a4paper]{elsart}

 \usepackage{graphics}
 \usepackage{graphicx}
 \usepackage{epsfig}
 \usepackage{amssymb}
 \usepackage{lineno}
 \usepackage{amsmath}
 \usepackage{epstopdf}
\usepackage{subfigure}
\usepackage{textcomp}
\usepackage{eurosym} 
\usepackage{url} 
\usepackage[T1]{fontenc}
\begin{document}

\begin{frontmatter}
\title{\bf Evidence for $\boldsymbol{\pi K}$-atoms with DIRAC}
\thanks[thanks1]{This publication is dedicated to the memory of Ludwig Tauscher.}
\author[o]{B. Adeva},
\author[l]{L. Afanasyev},
\author[r]{Y. Allkofer\corauthref{cor1}}, 
\author[r]{C. Amsler\corauthref{cor1}}, 
\ead{claude.amsler@cern.ch}
\author[f]{A. Anania},
\author[r]{A. Benelli}, 
\author[n]{V. Brekhovskikh},
\author[k]{G. Caragheorgheopol},
\author[b]{T. Cechak}, 
\author[j]{M. Chiba}, 
\author[n]{P. Chliapnikov},
\author[k]{C. Ciocarlan},
\author[k]{S. Constantinescu}, 
\author[e]{C. Curceanu}, 
\author[a]{C. Detraz},
\author[g]{D. Dreossi}, 
\author[a]{D. Drijard},
\author[l]{A. Dudarev},
\author[k]{M. Duma}, 
\author[k]{D. Dumitriu},
\author[o]{J.L. Fungueiri\~no}, 
\author[b]{J. Gerndt}, 
\author[n]{A. Gorin},
\author[l]{O. Gorchakov},
\author[l]{K. Gritsay},
\author[e]{C. Guaraldo},
\author[k]{M. Gugiu},
\author[a]{M. Hansroul},
\author[d]{Z. Hons},
\author[r]{S. Horikawa},
\author[e]{M. Iliescu},
\author[l]{V. Karpukhin}, 
\author[b]{J. Kluson},
\author[h]{M. Kobayashi},
\author[l]{V. Komarov}, 
\author[l]{V. Kruglov},
\author[l]{L. Kruglova},
\author[l]{A. Kulikov}, 
\author[l]{A. Kuptsov},
\author[n]{I. Kurochkin}, 
%\author[l]{K.-I. Kuroda},
\author[f]{A. Lamberto}, 
\author[e]{A. Lanaro}, 
\author[n]{V. Lapshin},
\author[c]{R. Lednicky}, 
\author[e]{P. Levi Sandri},
\author[o]{A. Lopez Aguera}, 
\author[e]{V. Lucherini},
\author[n]{I. Manuilov}, 
\author[o]{C. Mari\~nas},
\author[l]{L. Nemenov},
\author[l]{M. Nikitin}, 
\author[i]{K. Okada},
\author[l]{V. Olchevskii},
\author[k]{M. Pentia}, 
\author[g]{A. Penzo},
\author[o]{M. Pl\'o},
\author[f]{G.F. Rappazzo},
\author[r]{C. Regenfus},
\author[r]{J. Rochet},
\author[o]{A. Romero},
\author[n]{V. Ronjin}, 
\author[n]{A. Ryazantsev}, 
\author[n]{V. Rykalin},
\author[o]{J. Saborido}, 
\author[q]{J. Schacher},
\author[n]{A. Sidorov}, 
\author[c]{J. Smolik}, 
\author[h]{S. Sugimoto},
\author[i]{F. Takeutchi},
\author[l]{A. Tarasov},
\author[p]{L. Tauscher$^\star$}, 
\author[b]{T. Trojek},
\author[m]{S. Trusov},
\author[l]{V. Utkin},
\author[e,o]{O. V\'azquez Doce}, 
\author[b]{T. Vrba}, 
\author[m]{V. Yazkov},
\author[h]{Y. Yoshimura}, 
\author[l]{M. Zhabitsky},
\author[l]{P. Zrelov}

\newpage

\address[a]{CERN, Geneva, Switzerland}
\address[b]{Czech Technical University, Prague, Czech Republic}
\address[c]{Institute of Physics ACSR, Prague, Czech Republic}
\address[d]{Nuclear Physics Institute ASCR, Rez, Czech Republic}
\address[e]{INFN, Laboratori Nazionali di Frascati, Frascati, Italy}
\address[f]{INFN, Messina University, Messina, Italy}
\address[g]{INFN, Trieste University, Trieste, Italy}
\address[h]{KEK, Tsukuba, Japan}
\address[i]{Kyoto Sangyo University, Kyoto, Japan}
\address[j]{ Tokyo Metropolitan University, Japan}
\address[k]{IFIN-HH, National Institute for Physics and Nuclear Engineering,
           Bucharest, Romania}
\address[l]{JINR Dubna, Russia}
\address[m]{Skobeltsin Institute for Nuclear Physics of Moscow State
           University, Moscow, Russia}
\address[n]{IHEP Protvino, Russia}
\address[o]{IGFAE, Santiago de Compostela University, Spain}
\address[p]{Basel University, Switzerland}
\address[q]{Bern University, Switzerland}
\address[r]{Physik-Institut der Universit\"{a}t Z\"{u}rich, Switzerland}

\corauth[cor1]{Corresponding authors.}

\begin{abstract}
\noindent
We present  evidence for the first observation of electromagnetically bound $\pi^\pm K^\mp$-pairs ($\pi K$-atoms) with the DIRAC experiment at the CERN-PS. The $\pi K$-atoms are produced by the 24 GeV/c proton beam in a thin Pt-target and the $\pi^\pm$ and $K^\mp$-mesons from the atom dissociation are analyzed in a two-arm magnetic spectrometer.
The observed enhancement at low relative momentum corresponds to the production of 173 $\pm$ 54 $\pi K$-atoms.   The mean life of $\pi K$-atoms is related to the $s$-wave $\pi K$-scattering lengths, the measurement of which is the goal of the experiment. From these first data we derive a lower limit for the mean life of 0.8 fs at 90\% confidence level.
\end{abstract}
\pagebreak
\end{frontmatter}

% main text
\section{Introduction}
\label{introduction}
The study of electromagnetically bound hadronic pairs is an excellent method to probe  QCD at very low energy. Opposite charge long-lived hadrons such as pions and kaons can form hydrogen-like atoms bound by the Coulomb interaction. The strong interaction leads to a broadening of the atomic levels and dominates the lifetime of the atom. 

Pion-pion interaction at low energy, constrained by the 
approximate 2-flavour SU(2) (u,d) chiral symmetry, is the simplest and 
well understood hadron-hadron process \cite{WEIN66,LEUTW,COLA01}. The observation of $\pi^+\pi^-$-atoms ($A_{ 2\pi}$) was reported in ref. \cite{AFAN94} and a measurement of their mean life in ref. \cite{AFAN05}. 

The low energy interaction between the pion and the heavier (strange)
kaon is a  tool to study the more general 3-flavour SU(3)
(u,d,s) structure of hadronic interaction, which is not accessible in $\pi \pi$-interactions. 
A detailed
study of $\pi K$-interaction provides insights into a potential flavour (u,d,s) 
dependence of the crucial order parameter or quark condensate in Chiral
Perturbation Theory (ChPT) \cite{STER00}.

A measurement of the $\pi K$-atom ($A_{ \pi K}$) lifetime was proposed already in 1969
\cite{BILE69} to determine the difference $|a_{1/2}-a_{3/2}|$ of the $s$-wave $\pi K$-scattering lengths, where the indices {\small 1/2} and {\small 3/2} refer to the isospin of the $\pi K$-system. The $\pi^\pm K^\mp$-atom decays predominantly by strong interaction into the neutral meson pair $\pi^0 K^0$ ($\pi^0 \overline{K}^0$). The decay width of  the $\pi K$-atom in  the ground state is given by the relation \cite{BILE69, SCHW04}:
 
\begin{equation}
  \label{eq:julia0}
   \Gamma(A_{ \pi K})= \frac{1}{\tau_{1S}}=
   \frac{8}{9} \; \alpha^3 \; \mu^2 \; p^* \; |a_{1/2}-a_{3/2}|^2 \; (1+\delta).
\end{equation}
\noindent
$\tau_{1S}$ is the lifetime of the atom in the ground state, $\alpha$
the fine structure constant, $\mu$ the reduced $\pi^\pm K^\mp$ mass, and $p^*$ = 11.8 MeV/c the outgoing $K^0$ or
$\pi^0$ 3-momentum in the $\pi K$ center-of-mass system. The term $\delta$ $\simeq$ (4 $\pm$ 2) \%  \cite{SCHW04}  
accounts for corrections, due to isospin breaking  and the
quark mass difference $m_u - m_d$.
Hence a measurement of $\Gamma(A_{ \pi K})$ provides a value for the
scattering length $|a_{1/2}-a_{3/2}|$. The mean life of $\pi K$-atoms is predicted to be 3.7 $\pm$ 0.4 fs \cite{SCHW04}. 

The width  $\Gamma(A_{\pi K})$ can  also be determined from the  $s$-wave phase shifts obtained from $\pi K$-scattering, i.e. from the interaction of kaons with nucleons. The  $s$-wave phase shifts are, however,  poorly  known due to the absence of data below 600 MeV/c and, correspondingly, the uncertainties in $a_{1/2}$ and $a_{3/2}$ are  substantial. The overall interaction is attractive (attractive in the isospin {\small 1/2} and repulsive in the  {\small 3/2} state).  The Roy-Steiner equations  lead, with the available scattering data, to results  \cite{Buttiker} that are neither consistent with the most precise measurements \cite{Estabrook} nor with predictions from ChPT \cite{Bernard}.

The method used by DIRAC is to produce pions and kaons with a high energy proton beam impinging on a thin target. Pairs of oppositely charged mesons may interact and  form electromagneti\-cally bound systems. Their subsequent ionization in the production target leads to mesons emerging from the target with low relative momentum (thereafter called atomic pairs). The mean life of the $\pi K$-atom can then be calculated from the number of observed low-momentum pairs. This method was first proposed in 1985 \cite{NEME85} and  was successfully applied  to $\pi^+ \pi^-$-atoms  in Serpukhov on the U-70 synchrotron internal proton
beam  \cite{AFAN94},  and with DIRAC-I at the CERN-PS on beam line T8. Data from  DIRAC-I  lead to a mean life 
$\tau_{2 \pi} = (2.91 \pm _{0.62}^{0.49})$ fs for $\pi^+\pi^-$-atoms  \cite{AFAN05}.

Cross sections for $A_{ \pi K}$-production have been calculated in ref. \cite{GORC00}. In this paper we report on the observation of $\pi K$-atoms from the first data with DIRAC-II.
\section{Experimental setup}
 \label{experimental setup}

\begin{figure*}[htbp]%
\begin{center}
\includegraphics[width=0.85\textwidth]{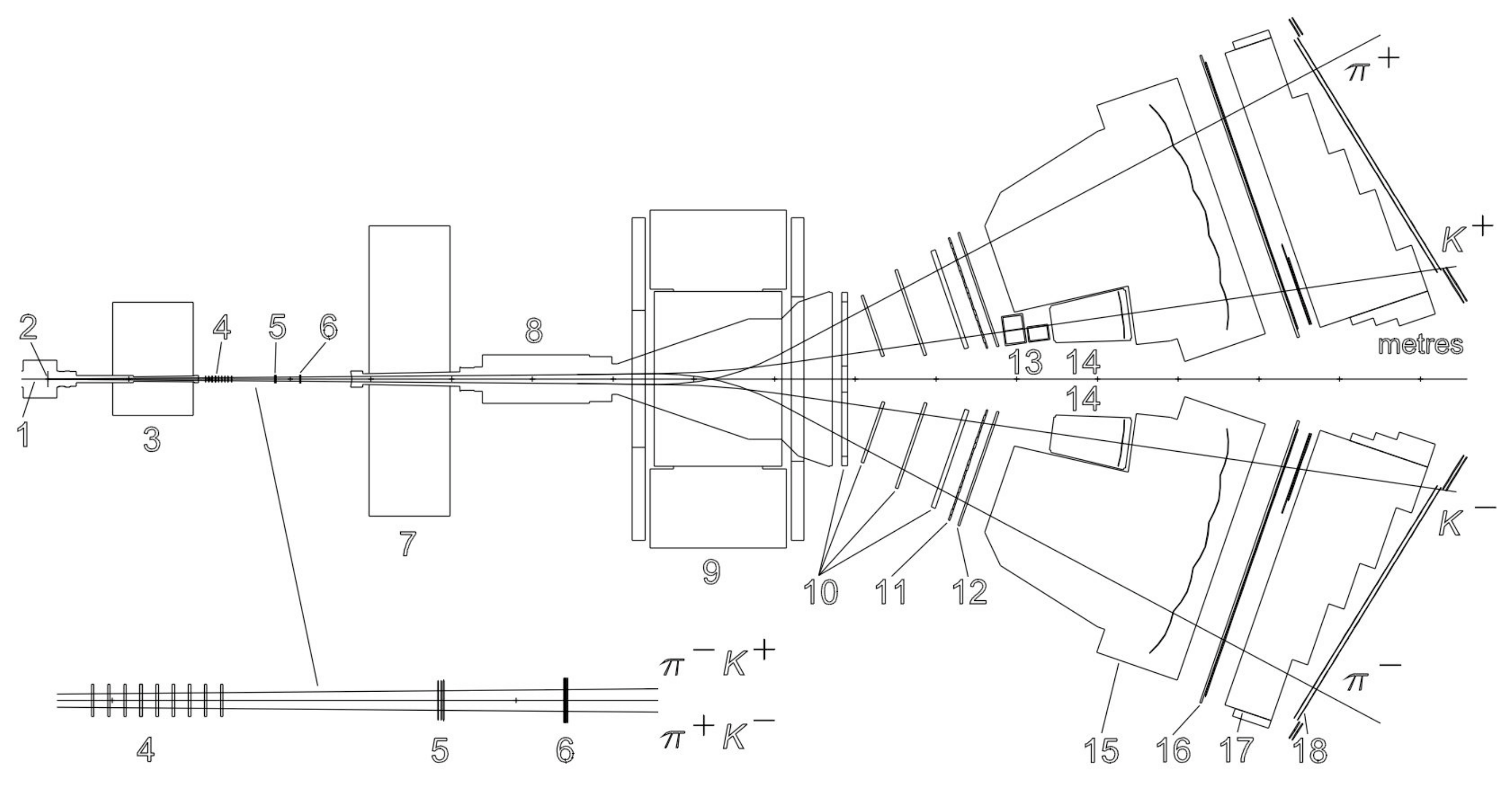}
\caption{ Sketch of the DIRAC-II spectrometer (top view). 1 -- proton beam, 2 -- production target, 3 -- shielding, 4 -- microdrift
chambers, 5 -- scintillation fiber detector, 6 -- ionization hodoscope,
7 -- shielding, 8 -- vacuum chamber, 9 -- dipole magnet, 10 -- drift chambers,
11 -- vertical hodoscope, 12 -- horizontal hodoscope, 13 -- aerogel
\v{C}erenkov modules, 14 -- heavy gas \v{C}erenkov detector, 15 -- 
N$_2$-\v{C}erenkov detector, 16 -- preshower detector, 17 -- absorber, 18 -- muon
scintillation hodoscope.  The  solid lines crossing the spectrometer arms correspond to typical $\pi^\pm$- and $K^\mp$-trajectories from the ionization of $\pi K$-atoms in the production target.}
\label{fig:a0smp}
\end{center}
\end{figure*}

Details on the initial  apparatus (DIRAC-I) used to study $\pi^+\pi^-$-atoms can be found in ref. \cite{ADEV03}. A sketch of the modified spectrometer (DIRAC-II) used 
to collect the $\pi K$ (and more $\pi\pi$) data  is shown in fig. 
\ref{fig:a0smp}. 
The 24 GeV/c proton beam {\bf (1)} from the CERN-PS impinges on a 26 $\mu$m $\rm Pt$-target {\bf (2)}.    The spill duration is 450 ms with an  average intensity  of $ 1.6 \times 10^{11}$ protons/spill.  The proton beam then passes through a vacuum pipe and is absorbed by the beam dump.  The secondary particles are collimated through two steel shielding blocks  {\bf (3)} and {\bf (7)}, upstream of  the microdrift chambers {\bf (4)} and downstream of the ionization hodoscope {\bf (6)}, respectively. They pass through a vacuum chamber {\bf (8)} and are bent by  the 1.65 T field of the dipole magnet {\bf (9)}. 
The  two-arm spectrometer is tilted upwards with respect to the proton beam by an angle of 5.7$^\circ$.  Positive particles are deflected into  the left arm, negative ones into the right arm. 

We now describe in more detail the upgrade  which was performed to search for and study $\pi K$-atoms \cite{ADEV04}.  The upstream detectors {\bf(4 -- 6)}, see fig. \ref{fig:a0smp}, were  either replaced or upgraded. However,  they were not yet fully operational during data taking, and were therefore not used in the analysis presented here. The tracking is performed by   drift chambers {\bf (10)} which have a spatial resolution of 85 $\mu$m. The vertical  hodoscope {\bf (11)} consisting of 20 scintillating slabs with a time resolution below 140 ps  is used for timing. The horizontal hodoscope {\bf (12)}, made of 16 horizontal scintillating slabs, is used for triggering by  selecting oppositely charged particles with a vertical displacement smaller than 75 mm. 

The   $\rm N_2$-\v{C}erenkov detector {\bf (15)} was already used previously to reject electrons and positrons. The  refractive index is $n=1.00029$ and the average number of photoelectrons $N_{pe}=16$ for  particles with velocity $\beta$ = 1. The   inner part of the original container had to be cut  to clear space for the two new \v{C}erenkov detectors needed for kaon identification. Since the momenta of the two mesons originating from  the breakup of the $\pi K$-atoms are very small in the center-of-mass system, they have similar velocities in the laboratory system, and hence kaons are less deflected than pions. Typical trajectories are shown in fig. \ref{fig:a0smp}.

The heavy gas $\rm C_4F_{10}$-\v{C}erenkov detectors in both arms  {\bf (14)} identify pions but do not respond to kaons nor (anti)protons  \cite{KUPT08}.  Four spherical and four flat mirrors each focus the light towards the phototubes.  The alignment of the mirrors was checked with a laser beam \cite{BREK08}. To keep a constant refractive index of $n=1.0014$, the gas has to be cleaned permanently with a complex recirculating system \cite{HORI08}. The ave\-rage number of   photoelectrons is 28 for particles with $\beta$ = 1. 

The aerogel \v{C}erenkov detector {\bf (13)} in the left arm identifies kaons and rejects protons \cite{ALLK07, ALLK08}. Such a detector is required only in the left arm since the contamination from antiprotons in the right arm is small due to their low production rate. The detector consists of three modules. Two modules  of 12$\ell$ each (refractive index $n=1.015$) cover the  relevant momentum range between 4 and 8 GeV/c.  The aerogel stacks are 42 cm high and are read out by two 5''-Photonis XP4570 photomutipliers with UV windows. The typical number of photoelectrons is 10 for $\beta$ = 1 particles. A third overlapping module with 13$\ell$ aerogel and $n=1.008$ covers the small angle region to reject protons with momenta above 5.3 GeV/c. 

The aerogel tiles are stacked pyramidally to increase the radiator thickness halfway between the photomutipliers and to compensate for light absorption. Due to the low light yield for the  $n=1.008$ module, and the strong UV-light absorption, we use a wavelength shifter. The aerogel tiles are sandwiched between Tetratex foils on which tetraphenylbutadiene (TPB) has been evaporated \cite{ALLK07}. The typical number of photoelectrons is 4 -- 8 for $\beta$ = 1 particles. 

The preshower detector {\bf (16)} provides additional electron/hadron separation in the offline analysis. A  lead converter, typically 10 -- 25 mm thick, is placed in front of a 10 mm thick scintillator.  An additional converter/scintillator is installed to compensate for the drop of efficiency  of the N$_2$-\v{C}erenkov detector at small angles, in the  region covered by the aerogel and heavy gas detectors. The iron absorber {\bf (17)} and the array of scintillation counters {\bf (18)} are used to suppress muons. 

Pairs such as $e^+e^-$, $\pi^+\pi^-$, $\pi^- K^+$ or $\pi^+ K^-$ are selected by a two-level trigger. For $\pi^-K^+$- and $\pi^+K^-$-candidates the first level trigger requires no signal in the heavy gas detector of the $K$-arm and in the N$_2$-\v{C}erenkov detectors. The two tracks have to cross the same (or one of the two adjacent) slab(s) in each of the horizontal hodoscopes. The two trajectories being asymmetric (see fig. \ref{fig:a0smp}), the kaon is required to cross  slabs of the vertical hodoscope located in front of the heavy gas detector, while the  associated pion has to fly at large angles in the opposite arm. The second level trigger \cite{AFAN02} uses raw hits from the drift chambers and, based on a lookup table, rejects pairs with high relative momentum. The accepted trigger rate is limited by the buffer memory of around 2000 events per spill. 

\section{Tracking and calibration} 
 \label{tracking and calibration}
As mentioned already, only  detectors downstream of the dipole magnet are used here for event reconstruction. The trajectories are determined by the drift chambers, the pattern recognition starting from the horizontal $x$-coordinate in the last plane and extrapolating back to the target. A straight line is first fitted through the hits and extended into the magnet yoke. A deflection algorithm \cite{Drijard} calculates the slope and $x$-coordinate of the track at the magnet entrance, following the magnetic field map. The trajectory is then extrapolated linearly to the target, with the constraint that the track origin has to coincide with the center of the beam spot. This determines the momentum of the particle.  For the vertical $y$-coordinate the straight line from the drift chamber information is extrapolated back to the center of the beam spot at the production target  \cite{ALLKthese,ALLK07a}.

The variable of interest in the following analysis is the relative momentum $Q$ of the $K^\pm\pi^\mp$-pairs in their center-of-mass systems. 
In the transverse plane, the resolution on the relative  momentum $Q_T$ (typically 3 MeV/c) is dominated by multiple scattering, while the resolution on the longitudinal component $Q_L$ ($<$ 1 MeV/c) is not affected \cite{ALLKthese,ALLK07b}. For  further analysis we use therefore only  $Q_L$. 

\begin{figure}[htb]
\begin{center}
\includegraphics[width=0.4\textwidth]{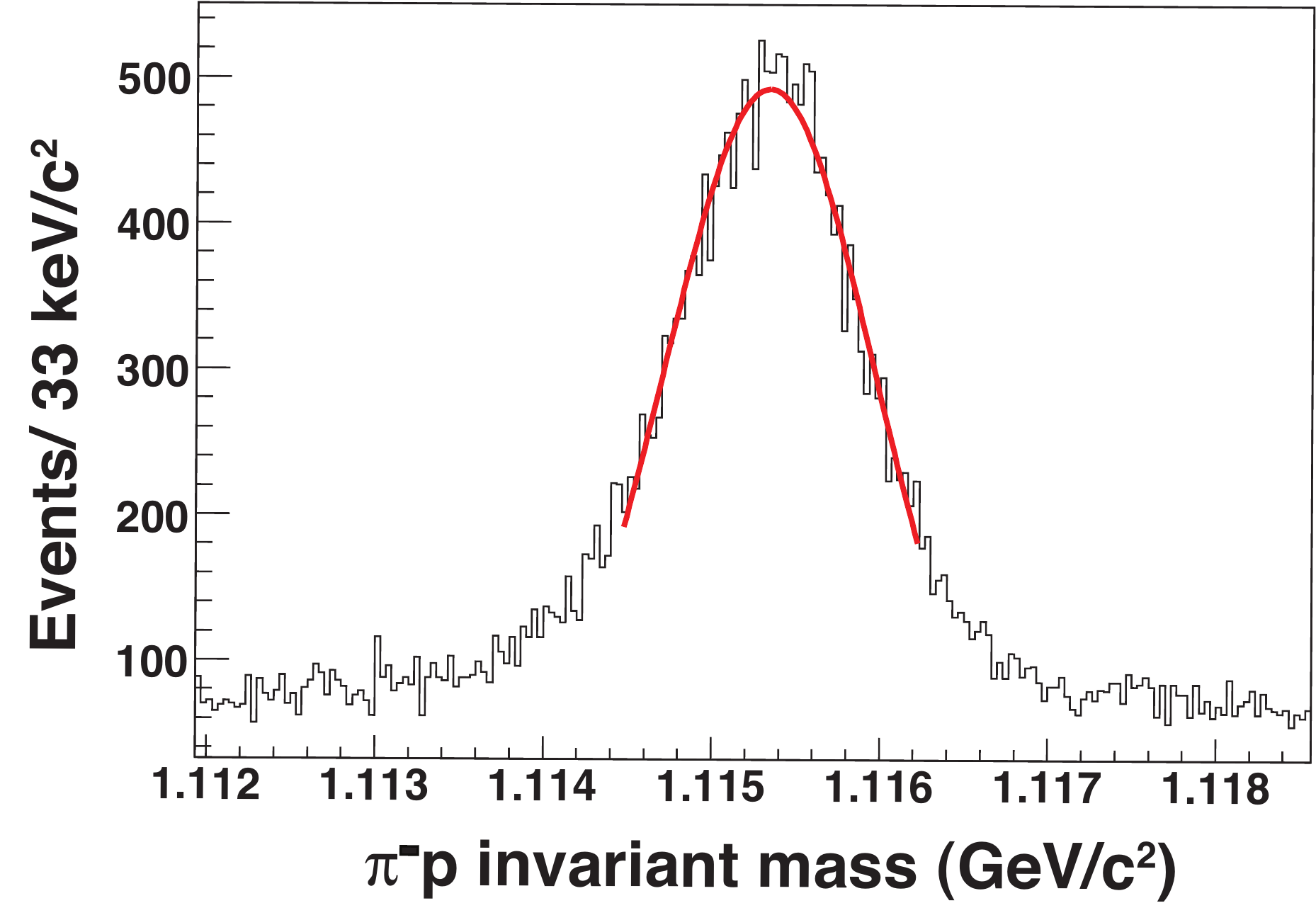}
\caption{{$\pi^-p$-mass distribution in the $\Lambda$-region. The   line shows the Gaussian fit.}}
\label{LAMBDA}
\end{center}
\end{figure}

The momentum calibration was cross-checked with tracks from  $\Lambda\to\pi^-p$ decays. Figure \ref{LAMBDA} shows the invariant $\pi^-p$-mass distribution. A Gaussian fit  is applied leading to a mass of  1115.35 $\pm$ 0.08 MeV/c$^2$ (statistical error) and a width of  $\sigma$ = 0.58 $\pm$ 0.01 MeV/c$^2$, dominated by  momentum resolution.

\begin{figure}[htb]
\begin{center}
\includegraphics[width=0.45\textwidth]{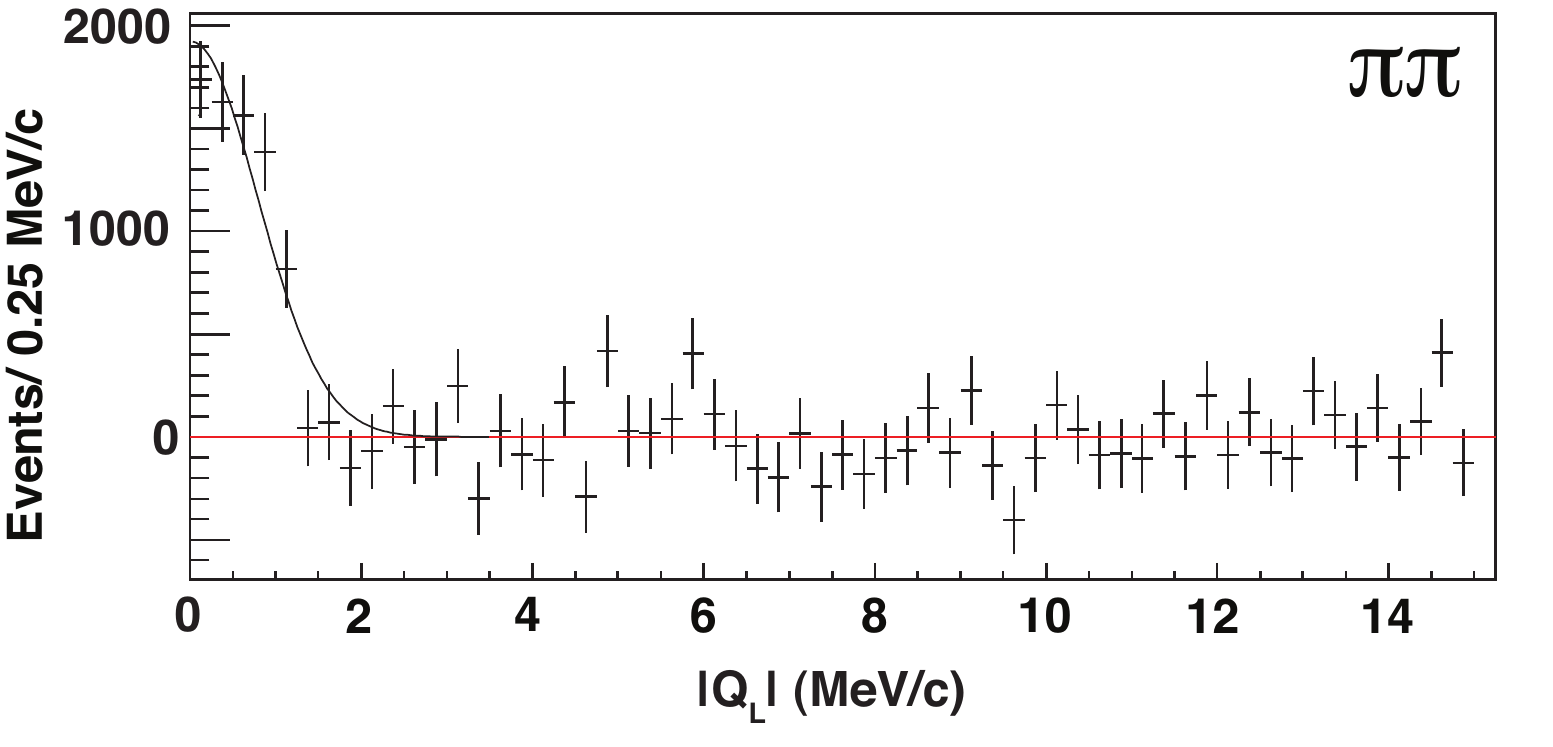}
\caption{{$Q_L$-distribution measured with DIRAC-II for  part of the $\pi^+\pi^-$ data after  background subtraction. The accumulation of events at low $Q_L$ is due 
to $\pi^+\pi^-$-atoms. The curve is a Gaussian fit to guide the eye.}}
\label{Pipiatoms}
\end{center}
\end{figure}

 We have verified that the new hardware and software were able to reproduce the signal from $\pi^+\pi^-$-atoms. Details of the procedure for $\pi^+\pi^-$ atoms can be found in ref. \cite{AFAN05,ALLKthese}. Figure~\ref{Pipiatoms} shows the momentum distribution after background subtraction. The enhancement at low $Q_L$  corresponds to $7098\pm533$ atomic $\pi^+\pi^-$-pairs.  Bound  $\pi^+\pi^-$-pairs are not expected above $|Q_L|=2$ MeV/c and indeed the distribution in fig.~\ref{Pipiatoms} is flat and compatible with zero, which validates the background subtraction method.
Note that the present data cannot be compared directly with those of ref. \cite{AFAN05} because we used here only the downstream detectors, and a different event selection was performed.

 \section{Data analysis}
 \label{data analysis and results}
 
\begin{figure*}[htb]
\begin{center}
\includegraphics[width=0.80\textwidth]{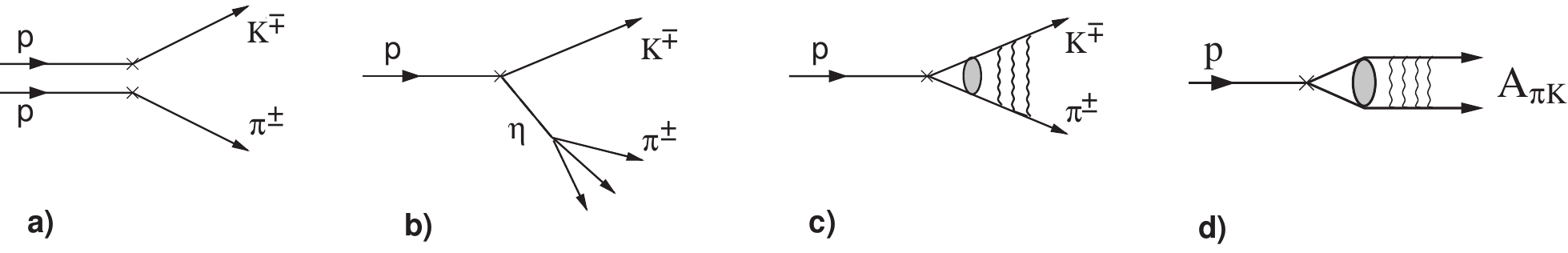}
\caption{{Production mechanisms of $\pi K$-pairs: a) accidental-pairs from two protons; b) non-Coulomb-pairs from  long-lived intermediate states  such as the $\eta$-meson; c) Coulomb-pairs from direct  production or from short-lived intermediate states; \newline d) $\pi K$-atoms.}}
\label{diagrampairs}
\end{center}
\end{figure*}

Figure~\ref{diagrampairs} shows the four mechanisms which contribute to the production of $\pi^\pm K^\mp$-pairs. Accidental pairs  are due to particles produced on different nucleons  (fig.~\ref{diagrampairs}a), non-Coulomb-pairs  are associated with the production of  long-lived intermediate states (fig.~\ref{diagrampairs}b).  On the other hand, $\pi^\pm K^\mp$-pairs which interact electromagnetically form correlated Coulomb-pairs  (fig.~\ref{diagrampairs}c), or  atomic bound states (fig.~\ref{diagrampairs}d). The $N^{A}$ atoms, while traveling through the target, can either decay, be (de)-excited  or break up  into $n^{A}$  $\pi^\pm K^\mp$-pairs which emerge from the target with very 
low relative momentum. 

We now describe the analysis steps \cite{ALLKthese}. For prompt pairs the time difference between the posi\-tive and negative spectrometer arm lies between --0.5 and 0.5 ns. Accidental pairs are first removed using the time information from the vertical hodoscopes.  Accidental pairs are also needed for  subsequent analysis and we select those pairs with a time difference between --12 and --6 ns. The choice of the negative sign avoids the contamination from slow protons.  The events have then to satisfy the following criteria:

\begin{itemize}
\item no electrons nor positrons,
\item no muons,
\item one drift chamber track per arm,
\item $|Q_L|<20$  MeV/c,
\item $Q_T<8$ MeV/c,
\item the momentum of the kaon lies between 4 GeV/c and  8 GeV/c,
\item the momentum of the pion  lies between 1.2 GeV/c and  2.1 GeV/c.
\end{itemize}

\noindent 
With the excellent time resolution of the vertical hodoscope  pions, kaons and protons  below 2.5 GeV/c can be separated by time-of-flight  \cite{ADEV02}. 
For the $\pi^- K^+$ analysis the  aerogel detector is used in addition to remove protons in the posi\-tive arm, while for the $\pi^+ K^-$ analysis,  the time difference between the negative and the  positive arm measured with the vertical hodoscope has to be negative in order  to remove protons faking  pions.  
Once the accidentals have been subtracted the prompt pairs ($N^{pr}$) are  composed of the following three types: atomic-pairs ($n^{A}$),   
Coulomb-pairs  ($N^{C}$), and non-Coulomb-pairs ($N^{nC}$). Therefore 
\begin{equation}
N^{pr}=n^A+N^{C}+N^{nC}.
\label{in the data sample}
\end{equation}

\begin{figure*}[htb]
\begin{minipage}[t]{0.50\textwidth}
\centering
\includegraphics[width=0.90\textwidth]{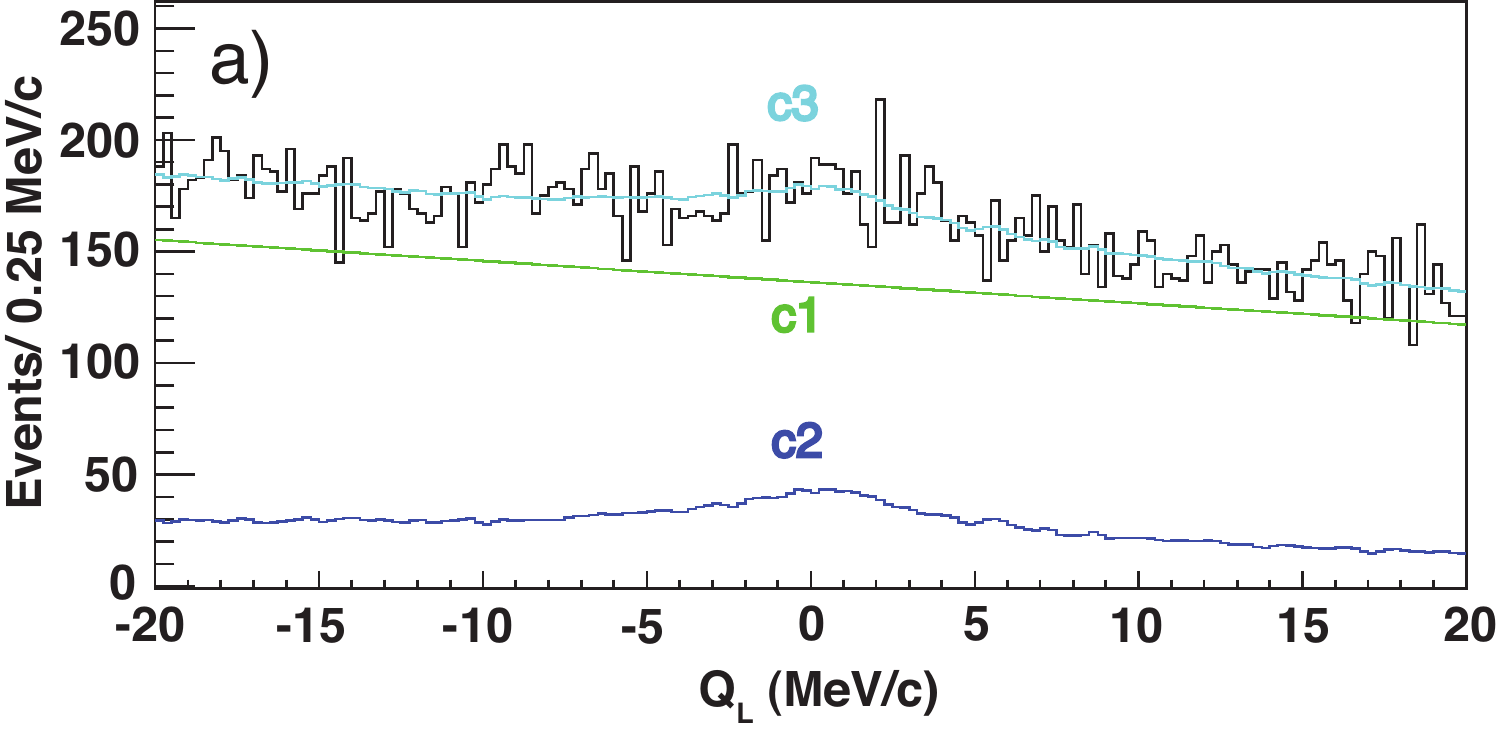}
\end{minipage}
\begin{minipage}[t]{0.50\textwidth}
\centering
\includegraphics[width=0.95\textwidth]{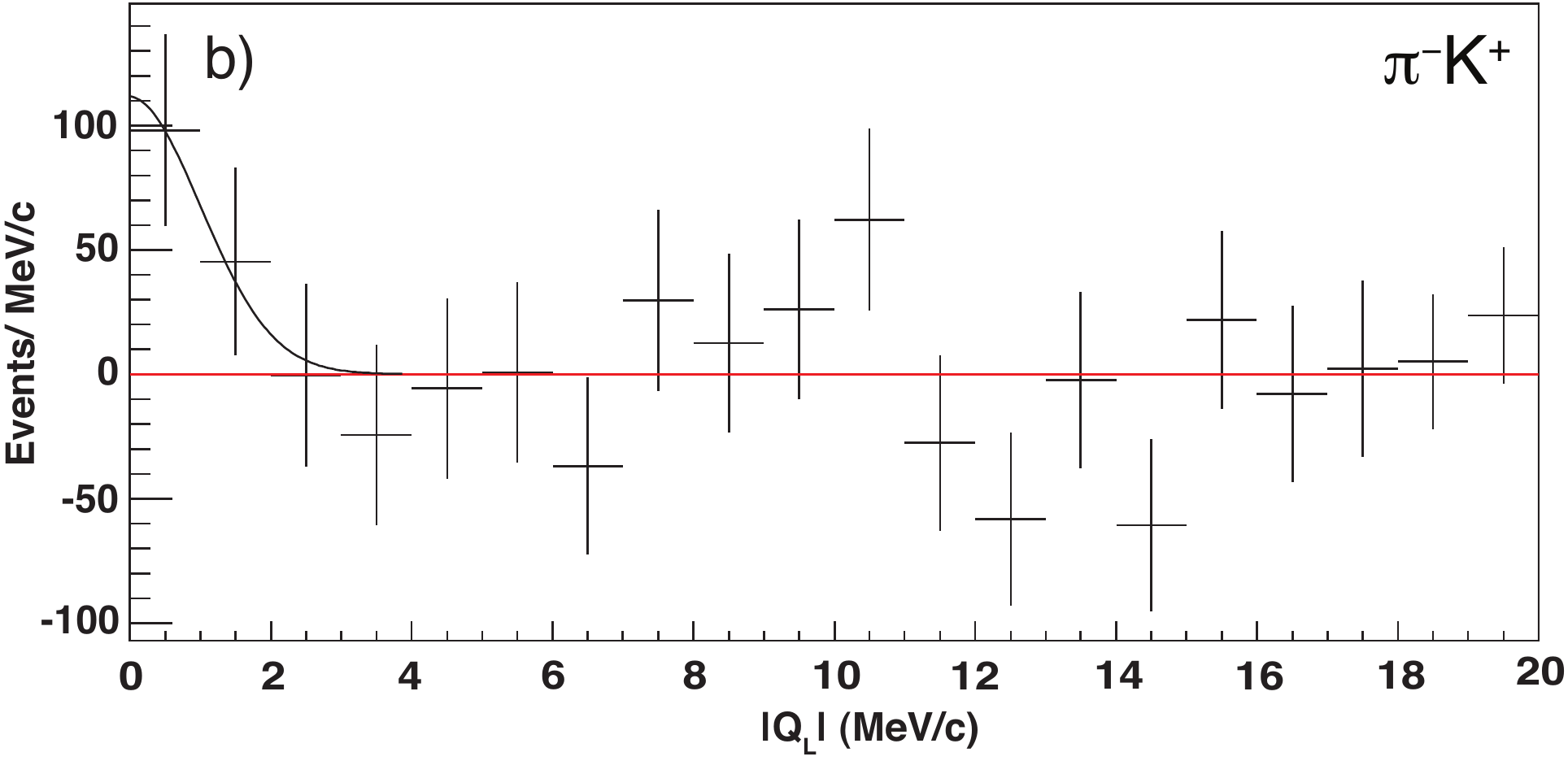}
\end{minipage}
\caption{{ a)  $Q_L$-distributions for the $\pi^- K^+$  data sample (26 $\mu$m Pt-target). The  histogram shows the data. The fitted  non-Coulomb-  and  Coulomb-pairs are in  green (c1) and blue (c2) respectively, together with the total background in turquoise (c3).
b) Residuals between data and fitted background. The solid line illustrates the distribution of atomic-pairs (see text).}}
\label{K+pi- fit}
\end{figure*}

However, the background arising from $\pi^+\pi^-$- and $\pi^- p$-pairs with misidentified particles must be considered since the  kaon flux is much lower than the pion  and proton fluxes.  Pions can be selected with the heavy gas detector in coincidence and protons with the aerogel detector in anticoincidence. We then determine  $Q_L$ by assigning to the pion (or the proton) the mass of a kaon.  For $\pi^+\pi^-$ events this incorrect mass assignment shifts the $Q_L$-distribution  by --150 MeV/c  \cite{ALLKthese} and  therefore does not overlap with the $Q_L$-distribution of $\pi K$ events. The contribution from $\pi^- p$-pairs,  non-Coulomb $\pi^- K^+$-pairs and accidentals have a similar linear $Q_L$-distribution. We assume that the background  due to Coulomb uncorrelated pairs  can be described by the $Q_L$-distribution of accidentals,  following a similar analysis for $\pi^+\pi^-$-atoms \cite{AFAN05}. The non-Coulomb background also includes the background from $\pi^-p$-pairs. Coulomb correlated pairs have to be simulated  \cite{SAK48,LED04}. 

To determine the contribution from  Coulomb- and non-Coulomb pairs we  select the momentum range $3 < |Q_L| < $ 20 MeV/c, where no atoms are expected.
The $Q_L$-distribution is
\begin{equation}
\frac{dN^{pr}}{dQ_L}=\beta\cdot\frac{dN^{C}}{dQ_L}+(N^{pr}-\beta)\cdot\frac{dN^{acc}}{dQ_L},
\label{in the data sample2}
\end{equation}
where  $dN^{C}/dQ_L$ and $dN^{acc}/dQ_L$  are the (normali\-zed) diffe\-rential probabilities for  Coulomb correlated or uncorrelated pairs, respectively.   The  fit variable $\beta$ is the corresponding number of correlated pairs. We choose a bin size of 0.25 MeV/c. The $\chi^2$-function to be minimized is 

\begin{eqnarray}
\chi^2 &=& \sum_{|i|=13}^{80}\left[\frac{\frac{dN^{pr}}{dQ_{L,i}}-\beta\cdot\frac{dN^{C}}{dQ_{L,i}}-(N^{pr}-\beta)\cdot\frac{dN^{acc}}{dQ_{L,i}}}{\sigma_i^{pr}}\right]^2,\nonumber\\
& &
\label{kicarre44}
\end{eqnarray}
\noindent
where $\sigma_i^{pr}$ are the corresponding statistical errors in the measured number of prompt pairs. Figure~\ref{K+pi- fit}a shows the MINUIT results  for Coulomb- and non-Coulomb contributions to $\pi^- K^+$ events. Since the shapes of both contributions are known,  one can extrapolate into the $|Q_L|<3$ MeV/c  signal region. The difference (residuals) between the data and the sum of both contributions is plotted in fig.~\ref{K+pi- fit}b. Above  $|Q_L| = 3$ MeV/c the residuals are consistent with zero, while the enhancement at low relative momentum is the first evidence for $\pi^- K^+$-atoms. 

A Gaussian distribution describes adequately the low momentum enhancement observed in the  $\pi^+\pi^-$ data \cite{ALLKthese}. To guide the eye we also apply a Gaussian fit here 
 (line in fig. \ref{K+pi- fit}b). The integral of the Gaussian distribution contains 147 $\pm$ 61  atomic $\pi^- K^+$-pairs.
 
  \begin{table}[htbp]
       \centering
      \begin{tabular}{c  c  r  r  r | r r}
      \hline 
      \multicolumn{5}{c|}{$\chi^2$-minimization} &   \multicolumn{2}{c}{Expected}\\

\hline
     Atom  &$\chi^2$/ndof & $\beta$\ \ \ \ & $N^C$\ \  & $n^A$\ \  & $N^A_e$ & $n^A_e$  \\
      \hline      
          
       $\pi^- K^+$\ &\ 0.92  \ & \  4215 \  & \  972 \   &  \ 143 \  & \  204\  &  108 \\
       &&\ $\pm$1008 \  & \ $\pm$233 \  &\  $\pm$53 \ &\  $\  \pm$ 59\  &  $\pm$ 31\\
         $\pi^+ K^-$\  &\  1.24 \ &\  1356 \  &\  164  \ &\  29 \ & \  74\ & 39\\
        &&\  $\pm$396 \  &\  $\pm$108 \  &\  $\pm$15 \  &\  \  $\pm$35\  &  $\pm$19\\
                    \hline
       \end{tabular}
       \vspace*{2mm}
       \caption{{Left: number $\beta$ of Coulomb-pairs  outside the signal region ($3 <  |Q_L| <  20$ MeV/c), number $N^C$ of Coulomb-pairs  extrapolated into the signal region ($|Q_L| <  3 $ MeV/c), and number $n^A$ of detected atomic-pairs from the residuals of the fit. Right: expected number of atoms $N^A_e$, calculated from the number of detected Coulomb-pairs, and expected number of atomic pairs $n^A_e$ using a breakup probability of 53\%. }}
      \label{final fit}
 \end{table}

A similar fit based on equ.  (\ref{kicarre44}) 
 is applied to $\pi^+ K^-$ events. However, the number of events is smaller, due to the  lower production cross section for negative kaons. The fit results are summarized in table~\ref{final fit}. The errors on the number of Coulomb-pairs $\beta$ are  the full (MINOS) errors, while the errors on  $n^A$  are given by the square roots of the measured bin contents, statistical fluctuations on the much larger Monte-Carlo sample being negligible.

\begin{figure}[htbp]
\begin{center}
\includegraphics[width=0.48\textwidth]{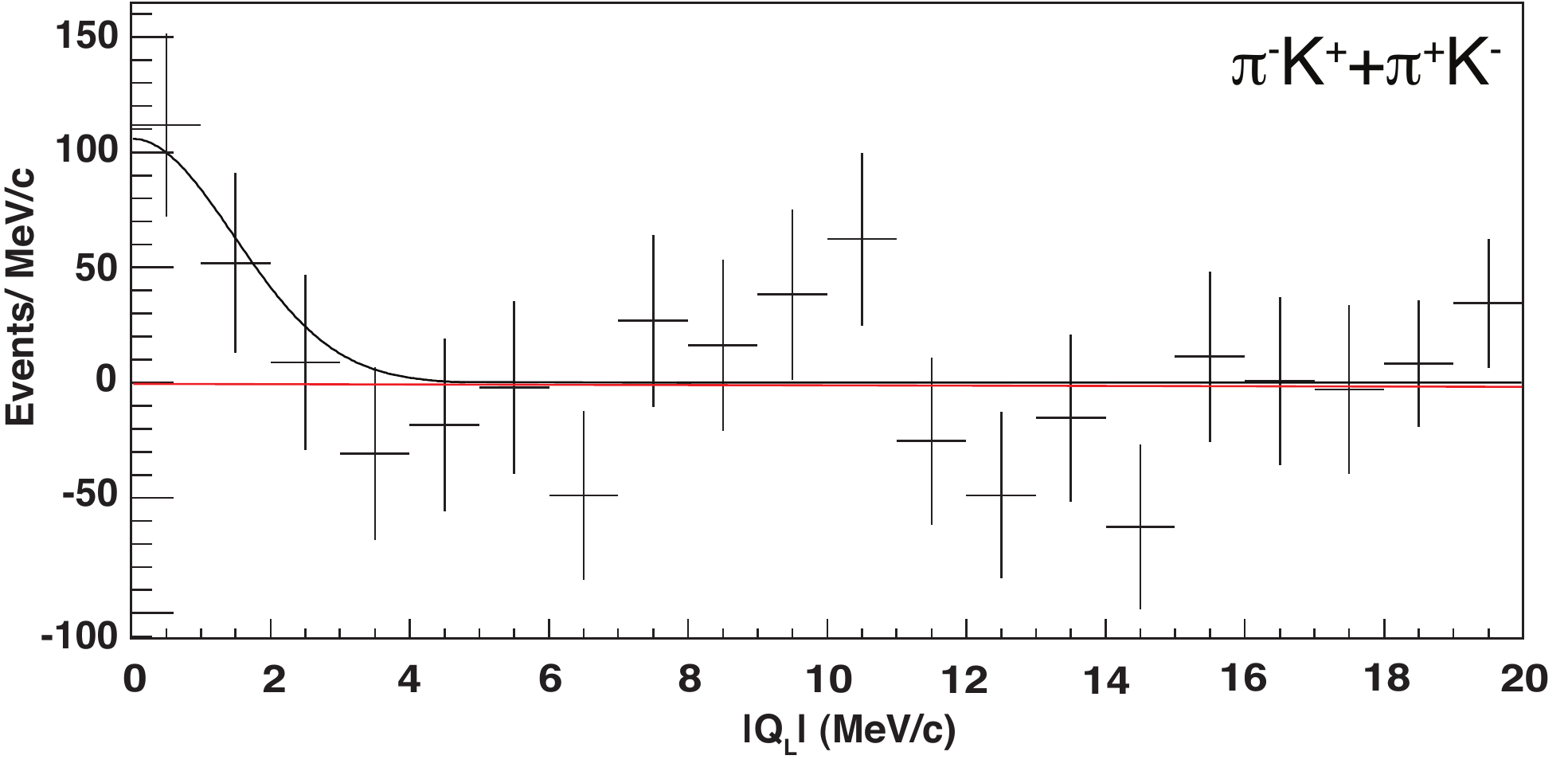}
\caption{Residuals between data and the fitted background  for  $\pi^- K^+$ and $\pi^+ K^-$.  A Gaussian fit has been applied  (solid line) to  illustrate the distribution of atomic-pairs.}
\label{totalrres}
\end{center}
\end{figure}

Figure~\ref{totalrres} shows the sum of the   $\pi^- K^+$ and $\pi^+ K^-$ residuals. We obtain
\begin{equation}
n^A(\pi^\pm K^\mp)=173\pm54
\end{equation}
detected atomic pairs with a statistical significance of 3.2$\sigma$. The systematic uncertainty is estimated to be around 5\%, much smaller than the statistical one.

\begin{figure}[htbp]
\begin{center}
\includegraphics[width=0.45\textwidth]{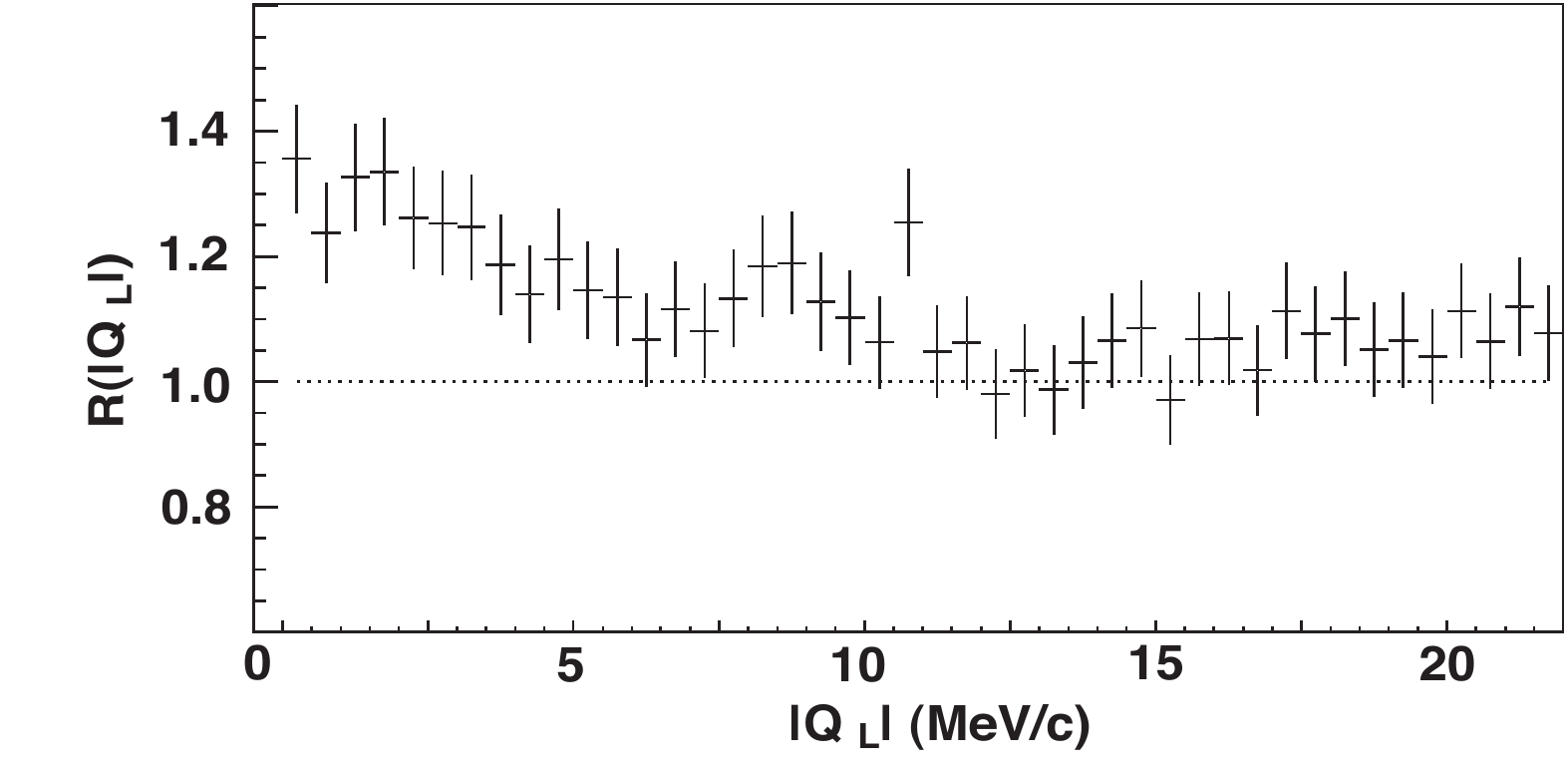}
\caption{{Correlation function $R$ as a function of $|Q_L|$ for $\pi^- K^+$-pairs. The deviation  from the horizontal  dotted line proves the existence of Coulomb correlated $\pi^-K^+$-pairs.}}
\label{Corr}
\end{center}
\end{figure}

The evidence for the observation of $\pi K$-atoms is strengthened by the observation of correlated Coulomb-pairs which, a fortiori, implies that atoms have also been produced. This can be seen as follows, without involving simulation: non-Coulomb pairs have a similar $Q_L$-distribution as accidentals. Hence dividing the normalized distribution for prompt pairs by the one for accidentals one obtains the correlation function $R$ describing Coulomb-pairs. The function $R$, shown in fig. \ref{Corr} for $\pi^- K^+$ as a function of $|Q_L|$, is clearly increasing with decreasing momentum, proving that Coulomb-pairs have been observed.  In the signal region ($|Q_L|<3$ MeV/c) one obtains   858 $\pm$ 247  Coulomb-pairs from  the data in fig. \ref{Corr}, without  resorting to Monte-Carlo.  The same procedure can be applied to $\pi^+ K^-$ events, leading to   313 $ \pm $ 148 Coulomb-pairs \cite{ALLKthese}. 
 
The ratio $k$ of the  number of produced atoms to the  number of Coulomb-pairs with small relative momenta has been calculated: $k$ = 0.62 \cite{NEME85,AFAN99}. However, one needs to take into account the acceptance of the apparatus and the cuts applied in the analysis. By Monte-Carlo simulation \cite{GorchMC} we determine the ratio  $k_{exp}$ = 0.24 between the number of  atoms produced within acceptance and the number of detected Coulomb-pairs below $|Q_L|$ = 3 MeV/c (and below $Q_T$ = 8 MeV/c).  This then leads to the expected number $N^A_e$ of atoms, 204 $\pm$ 59 for $\pi^+ K^-$, and 74 $\pm$ 35  for $\pi^- K^+$ (table \ref{final fit}, right). The uncertainty on $k_{exp}$ is negligible.

\begin{figure}[htbp]
\begin{center}
\includegraphics[width=0.30\textwidth]{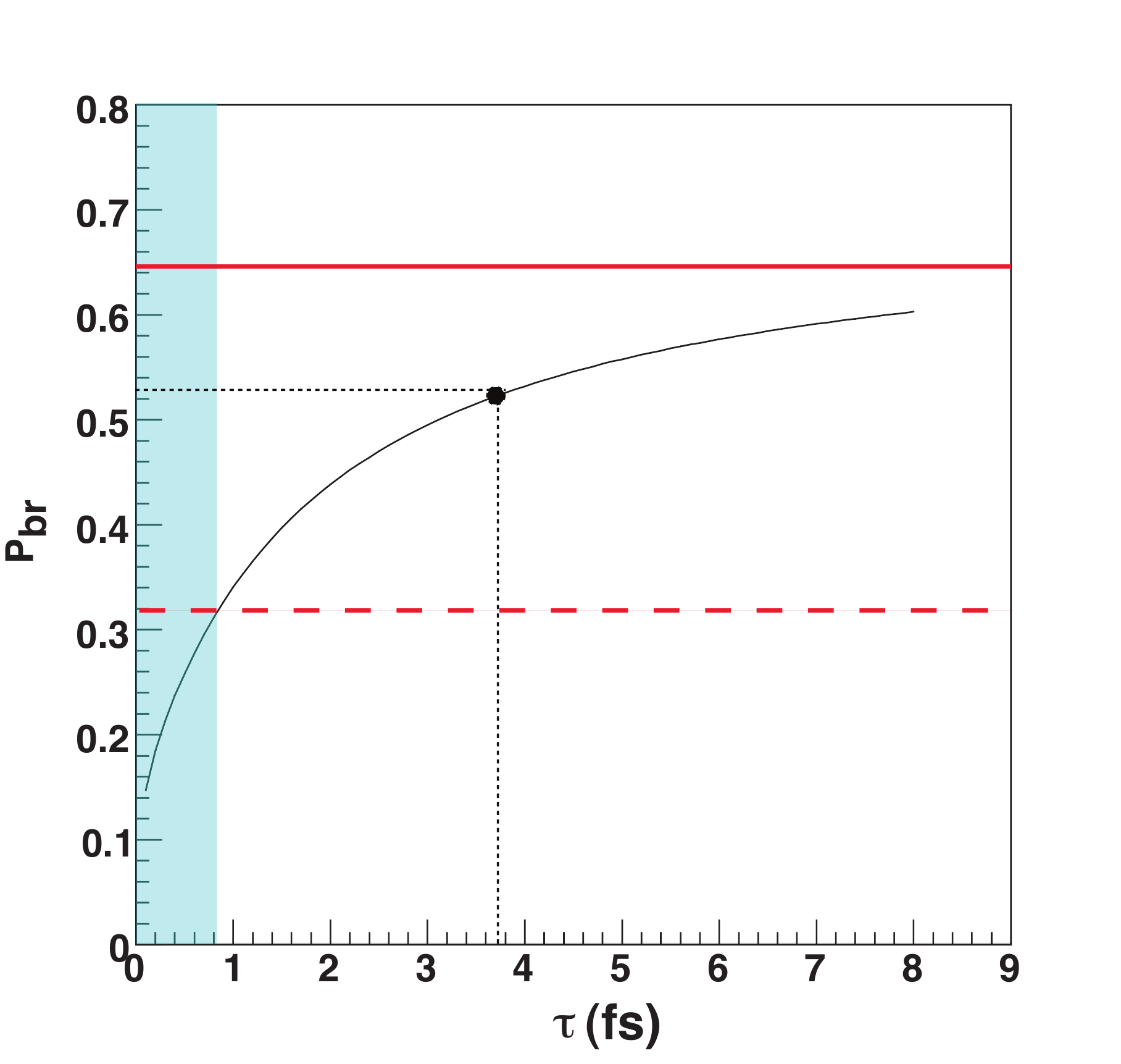}
\caption{Breakup probability $P_{br}$ for the $26\ \mu$m Pt-target as a function of mean life of $\pi K$-atoms in the 1s-state.  The  horizontal solid line is the measured breakup probability and the horizontal dashed line the  $1.28\sigma$ lower bound corresponding to a lower limit of 0.8 fs for the mean life. The excluded area (90\% confidence level)  is shown in turquoise. The horizontal  dotted line gives the theoretical prediction \cite{SCHW04}. }
\label{lifetime}
\end{center}
\end{figure}

The breakup probability 
\begin{equation}
P_{br}=\frac{n^A_e}{N^A_e}
\label{breakup}
\end{equation} 
relates the number of atoms to the number of atomic pairs. 
A calculation of the breakup probability as a function of mean life (fig.~\ref{lifetime}) has been performed using the Born approximation \cite{ADEV04}. For the predicted mean life of 3.7 $\pm$ 0.4 fs \cite{SCHW04} the corres\-ponding breakup probability $P_{br}$ is 53\% (dotted line in fig.~\ref{lifetime}). From equ. (\ref{breakup}) we then find the number $n^A_e$ of expected pairs  given in table  \ref{final fit}, right. These numbers are in good agreement with the number $n^A$ of observed atomic-pairs (fifth column in table  \ref{final fit}).    
 
Conversely, we use the number of observed atomic-pairs $n^A$ from  the $\chi^2$-minimization  and the number $N^C$ of Coulomb-pairs below $|Q_L|<$ 3 MeV/c (table  \ref{final fit}) to calculate the breakup probability  $P_{br}$ from equ. (\ref{breakup}) with $N^A= k_{exp} \cdot N^C$. The result for $\pi^\pm K^\mp$ ($P_{br}$ = 64 $\pm$ 25 \%) is shown by the horizontal solid line in fig.~\ref{lifetime}. This leads to a lower limit for the mean life of $\pi K$-atoms of $\tau_{1S} $ = 0.8  fs at a  confidence level of 90\%. This result can be translated into an upper limit  $|a_{1/2}-a_{3/2}|$ $<$ 0.58 $m_\pi^{-1}$ at 90\% confidence level, in agreement with predictions \cite{SCHW04,Buttiker}.

\section{Conclusions}
 \label{conclusion}

We have presented the first evidence for the production of $\pi K$-atoms by detecting $173 \pm 54$ atomic-pairs. The evidence is strengthened by the observation of correlated $\pi K$ (continuum) Coulomb-pairs from which the number of bound states (atoms) is predicted and found to be in  agreement with observation.  A lower limit on the mean life of 0.8 fs  is established   with a confidence level of $90\%$. We note that the choice of Pt as production target was driven by the high breakup probabi\-lity facilitating the observation of $\pi K$-atoms. Data are now being collected for a more accurate measurement of the lifetime with e.g. a 98 $\mu$m Ni-target, for which the breakup probability is lower ($\sim$35\% according to ref. \cite{ADEV04}) but still rapidly rising around the predicted mean life of  3.7  fs. The ultimate goal of the experiment is to measure the lifetime of $\pi K$-atoms with a precision of about 20\% \cite{ADEV04}.
 
 \section*{Acknowledgements}
 \label{acknowledgement}
We are grateful to the CERN-PS crew, whose efforts permitted us to take advantage of a high qua\-lity beam.  
This work was supported by CERN, the Grant Agency of the Czech 
Republic, 
the Istituto 
Nazionale di Fisica Nucleare (Italy), the Grant-in-Aid 
for Scientific Research from the Japan Society for the 
Promotion of Science, 
the Ministry 
of Education and Research
(Romania), the Ministery of Industry, 
Science and Technologies of the Russian Federation and 
the Russian Foundation for Basic Research,
the Direcci\'on Xeral de Investigaci\'on, Desenvolvemento e Innovaci\'on,
Xunta de Galicia (Spain), and the Swiss National 
Science Foundation. 

This publication summarizes the PhD thesis  of Y.~Allkofer.

\end{document}